\documentstyle[12pt,aaspp4]{article}
\def\yrm1{yr${}^{-1}$}
\def\sun{\odot}
\def\msun{$M_{\sun}$~}

\def\msunend{$M_{\sun}$}

\def\nco{${^{14}N}(e^-,\gamma){^{14}C}(\alpha,\gamma){^{18}O}$~}
\def\3a{3$\alpha$~}
\def\n14{${^{14}}$N}
\def\c14{${^{14}}$C}
\def\o18{${^{18}}$O}
\def\dash{------}
\begin{document}
\title{The Influence of ${^{14}N}({e^-},\nu){^{14}C}(\alpha,\gamma){^{18}O}$ 
reaction on the He-Ignition in Degenerate Physical Conditions}
\author{Luciano Piersanti  \altaffilmark{1}}
\affil{Dipartimento di Fisica dell'Universit\`a degli Studi di Napoli
``Federico II'', Mostra d'Oltremare, pad. 20, 80125, Napoli, Italy; 
lpiersanti@astrte.te.astro.it}

\author{Santi Cassisi}
\affil{Osservatorio Astronomico di Teramo, Via M.Maggini 47, 
64100 Teramo, Italy; cassisi@astrte.te.astro.it} 

\and

\author{Amedeo Tornamb\'e \altaffilmark{2}}
\affil{Osservatorio Astronomico di Teramo, Via M.Maggini 47, 
64100 Teramo, Italy; tornambe@astrte.te.astro.it} 

\noindent
\altaffilmark{1}{Osservatorio Astronomico di Teramo, Via M.Maggini 
47, 64100 Teramo, Italy} 
\altaffilmark{2}{Dipartimento di Fisica, Universit\`a de L'Aquila,
Via Vetoio, 67100 L'Aquila, Italy}

\begin{abstract}
The importance of \nco (NCO) chain on the onset of the He-flash in degenerate physical 
conditions has been reevaluated. We find that low-mass, metal-rich (Z $\ge$ 0.001) 
structures climbing the Red Giant Branch do never attain the physical conditions 
suitable for the onset of this chain, while at lower metallicities the energy 
contribution provided by NCO reaction is too low to affect the onset of the central 
He-flash. 

At the same time, our evolutionary models suggest that for a Carbon-Oxygen White Dwarf 
of mass $M_{WD}=0.6$\msun accreting He-rich matter, directly or as a by-product of an 
overlying H-burning shell, at rates suitable for a dynamical He-flash, the NCO energy 
contribution is not able to keep hot enough the He-shell and in turn to avoid the 
occurrence of a strong electron degeneracy and the ensuing final explosion.
\end{abstract}

{\em Subject headings}: nucleosynthesis - stars: accretion - supernovae: general 
- white dwarf

\section{Introduction}

More than 25 years ago Mitalas (1974) firstly suggested that the \nco (NCO) reaction 
could play an important role on the onset of the He-flash in the He core of a Red 
Giant (RG) star. In fact in high degenerate physical conditions, when the electron 
Fermi energy becomes of the order of the energy threshold for electron capture on \n14 
($\sim$ 156 keV), a significant amount of \c14 can be produced. It is important to 
notice that \c14 is unstable for $\beta^-$ decay, so that only for densities greater 
than a critical value ($\rho_{th}=\rho Y_e \sim 10^6$ g/cm$^3$, where $\rho$ is the 
mass density and $Y_e$ is the number of electrons per baryon) \c14 lives long enough 
to undergo an $\alpha$-capture. The latter reaction heats up the inner zones of the He 
core in such a way that the physical conditions suitable for the ignition of He-burning 
are attained before when compared with structures in which the NCO energy contribution 
is neglected. 
This implies that the He core mass at the onset of the central He-flash could be 
reduced and, as a consequence, the luminosity of both tip of RGB and Horizontal Branch 
(HB) could be fainter. This notwithstanding the main conclusion reached by Mitalas is 
that NCO reaction does not affect at all the evolution of a RG Population II star 
because the typical physical conditions of the He core do not allow the onset of 
electron capture.

However, Kaminisi, Arai \& Yoshinaga (1975) pointed out that Mitalas underestimated 
the cross section of the $\alpha$-capture on \c14 by a factor $10^6$. Moreover, 
Kaminisi \& Arai (1975) compared the energy released by the \nco chain and by \3a 
reaction respectively and concluded that NCO chain dominates over the \3a reaction for 
central temperatures and densities typical of a low metallicity RG star.

A new evaluation of the role played by NCO chain has been made by Spulak (1980). He 
pointed out that the onset of this reaction critically depends on the density so that 
its energy contribution is strongly concentrated toward the center of the He core in a 
RG structure, while the core He-flash driven by \3a occurs at the mass point where the 
temperature is maximum, which is off-center and approximately located at 
$M_{T-max}\sim 0.15$\msun. By accounting for the energy balance between the NCO 
chain, neutrino losses and \3a reaction, Spulak concluded that NCO chain is not able 
to trigger the onset of the He-flash because in a typical low metallicity RG star the 
central density becomes high enough to trigger the electron capture when the \3a 
reaction has already ignited off center and its energy contribution is quite relevant.

In 1984 Hashimoto, Nomoto, Arai \& Kaminisi (Hashimoto et al. 1984) recomputed the 
cross section for the electron capture on \n14 and once again they concluded that NCO 
reaction dominates over \3a reaction in the ignition of the He-flash. They also pointed 
out that the key role played by this reaction strongly depends on the chemical 
composition of the He core when the central density exceeds the density threshold for 
the electron capture. Using this updated cross section, Hashimoto et al. (1986) showed 
that NCO reaction plays an important role in the onset of the He-flash in He WDs 
accreting Helium rich matter directly or as a by-product of an overlying H-burning 
shell. Moreover, these authors pointed out that NCO reaction also affects the evolution 
of a CO WD accreting He rich matter directly or as a by-product of H-burning. In fact, 
for low accretion rates (i.e. smaller than $4\times 10^{-8}$ \msun \yrm1) the energy 
contribution provided by NCO reaction heats up the He layers, thus preventing them to 
undergo a violent dynamical He-flash and causing an increase in the mass of the CO core 
up to the Chandrasekhar mass limit. 

For this reason, Woosley \& Weaver (1994) included NCO chain in their nuclear network 
to construct the pre-supernova model of cool CO WDs accreting He rich matter. Their 
calculations show that NCO reaction does not affect the final outcome of the process 
(He detonation), since the main effect is a slight decrease in the mass of the He layer 
at the onset of the flash. 

More recently Piersanti et al. (1999) have included the NCO chain in the computation of 
the evolution of a low mass CO WD accreting He rich matter both directly and as a 
by-product of H-burning shell at rates suitable for the occurrence of a He detonation. 
They found that the differences produced by NCO chain are quite negligible and pointed 
out that this result is due to the fact that their initial WD model is an evolutionary 
one, obtained evolving an intermediate mass star with a moderate mass loss from the 
Zero Age Main Sequence down to the cooling sequence. In such a model there is no \n14 
available for NCO reaction at the physical base of the He-shell where the He-flash is 
ignited. 

In order to improve our knowledge on the role played by NCO chain in the framework of 
stellar evolution we have included this chain in our nuclear network, and performed 
several numerical experiments on low mass stars climbing the RGB, on central He-burning 
models, as well as on low mass CO WDs accreting He-rich matter. Moreover, we have also 
analyzed extensively the dependence of the NCO energy contribution on stellar 
metallicity. 

The plane of this paper is the following: 
in \S 2 we discuss the numerical techniques, the physical assumptions together with 
the input physics; 
in \S 3 we present the results concerning the evolution along the RGB, while 
in \S 4 we discuss the physical and chemical properties of CO WDs and explore the 
thermal behavior of a CO WD accreting He-rich matter.
In \S 5 a brief discussion and the conclusions close the paper.

\section{Physical Inputs}

All the models have been calculated with an update version of the FRANEC code (Chieffi 
\& Straniero 1989), the main differences have been discussed in Cassisi et al. (1998). 
As far as the equation of state (EOS) is concerned, we adopted the EOS computed by 
Straniero (1988). 
For low-temperature (T $< $12000 K) opacities we adopted the 
tables provided by Alexander \& Ferguson (1994) while for higher temperatures
the tables provided by Huebner et al. (1977) has been used. During the He 
burning phase we adopted a nuclear network which includes CO chain 
(${^{12}C}(\alpha,\gamma){^{16}O}(\alpha,\gamma){^{20}Ne}$), NO chain 
(${^{14}N}(\alpha,\gamma){^{18}O}(\alpha,\gamma){^{22}Ne}$) and NCO chain 
(${^{14}N}({e^-},\nu){^{14}C}(\alpha,\gamma){^{18}O}(\alpha,\gamma){^{22}Ne}$).
All the cross sections are from Caughlan \& Fowler (1988) with the exception of 
${^{12}C}(\alpha,\gamma){^{16}O}$ for which we adopted the cross section provided by 
Caughlan  et al. (1985, for a detailed discussion on the current uncertainty on this 
reaction see Bono et al. 
2000). The electron capture on \n14 was implemented according to Hashimoto et al. 
(1986). The evolution during the central He burning phase has been computed by taking 
into account semiconvection but by neglecting breathing pulses (Caputo et al. 1989).
The accretion process has been computed by assuming that the accreted matter has the 
same specific entropy of the matter at the surface of the WD (Limongi \& Tornamb\'e 
1991). For the heavy elements we assumed a solar scaled distribution (Grevesse 1991). 

\section{The evolution along the RGB}

We have computed three different sets of models at different metallicities, - Z=0.02, 
0.001, 0.0001 - from the pre-Main Sequence phase up to the ignition of the central 
He-burning. For each set of models we adopted three different evolving masses, namely 
M=0.6, 0.7 and 0.8 \msunend. Fig. 1 shows the $\rho - T$ plane for selected structures 
at various luminosity levels along the RGB up to the onset of the He-flash (the top 
curve in each panel). The dashed lines display the evolution of the central conditions 
of the models, while the heavy solid lines on the right of the tracks mark the region 
in which e-captures occur. Data plotted in this figure show that at higher metallicities 
(Z=0.02 and 0.001) the evolving stars do never attain densities high enough for the 
occurrence of e-captures. In these structures the NCO reaction does not become active 
at all, and thus it does not provide any contribution to the He-burning. 
On the contrary, in metal-poor structures (Z=0.0001) the density of the central regions 
becomes higher than the critical density for the onset of the NCO chain. Therefore its 
energy contribution might play a role in the He-burning ignition. 

\placefigure{fig1}     

To assess in more detail the effect of the NCO reaction in metal-poor structures, 
Fig. 2 shows the evolution in the $\rho - T$ plane of the central conditions for the 
0.6 \msun models which include (dotted line) or neglect (solid line) the NCO chain in 
the nuclear network. As soon as the central density exceeds the critical value 
$\rho_{th}$ ($\log (L/L_{\odot})=3.259$, $\log(T_e)=3.643$), the electron 
capture becomes active and produces \c14; the subsequent 
$\alpha$-capture on \c14 releases some amount of energy that heats up the center. 
However the abundance of \n14 nuclei is very low ($X_{^{14}N} < 10^{-4}$ by mass) and, 
in addition, the energy lost via neutrino emission are always active, and therefore
the center is barely heated. The top left panel of Fig. 1 clearly shows that, only the 
innermost zone of the He-core has a density larger than the critical value, and 
therefore the point where the He-burning is ignited via \3a reaction remains completely 
unaffected. This implies that at the onset of the He-flash - defined according to 
Sweigart \& Gross (1978) as the time when the energy production in the inner regions 
is greater than the neutrino energy losses - the He-core mass, and in turn the 
luminosity of the star are the same compared to the case in which NCO chain is not 
accounted for (see Fig. 3). 

\placefigure{fig2}

\placefigure{fig3}

\placefigure{fig4}

\placefigure{fig5}

\section{CO WDs Accreting Mass}

\subsection{The Chemical and Physical Properties of CO WDs}

To investigate the physical and chemical properties of CO WDs we computed a set of 
models from the central He burning phase down to the cooling sequence for different 
metallicities. Figures 4-5 show temperature and density profiles (right panels) 
together with the chemical abundances in the most external layers (left panels) for 
each model. Data plotted in the left panels show that the abundance of \n14  is equal 
to zero at the physical base of the He-shell, defined as the zone where the He 
abundance is Y=0.5 by mass. This is due to the fact that during the previous evolution 
(He-shell-burning phase) \n14 has been transformed into ${^{22}Ne}$ via NO chain. In 
fact, the ${^{14}N}(\alpha,\gamma){^{18}O}$ reaction becomes active at a temperature 
of the order of $8\times 10^7$ K with a mild dependence on density. As a consequence, 
the innermost zone of the He-shell is enriched in ${^{22}Ne}$ and deprived in \n14 .

\subsection{Accretion Experiments}

As already mentioned (see \S1), Woosley \& Weaver (1994) included NCO reaction in their 
nuclear network to compute the evolution of a low-mass CO WD accreting He-rich matter. 
By comparing their results with those of Limongi \& Tornamb\'e (1991), it is evident 
that for accretion rates suitable for the occurrence of a hydrodynamical event 
($\dot{M} < 4\times 10^{-8}$ \msun\yrm1), the energy contribution provided by NCO chain 
marginally affects the evolution of the accreting models. In fact, the final outcome 
remains an explosion and the only change is a slight reduction in the mass extension 
of the He layer at the onset of the He-flash. This result is confirmed by Piersanti et 
al. (1999) who pointed out that if the initial model of the CO WD is an evolutionary
one, then the influence of the NCO chain is negligible due to the lack of \n14 at the 
bottom of the He-shell. The same occurrence takes place over the thick He-layer, 
surrounding the CO core, where the Helium abundance has not been modified by \3a 
reaction, but the NO reaction has been active. In addition, these authors noticed that, 
for a fixed chemical  composition of the accreted matter, the energy contribution of 
NCO chain depends on the accretion rate, however the final outcome remaining unchanged. 
In fact, a decrease in the accretion rate causes an increase in the He-layer mass at 
the onset of the He-flash, and an increase in the ignition density as well. 
Therefore, the \c14 shell, defined as the point where the NCO reaction energy release 
 is at maximum, 
can move outward (in mass) detaching more and more from the base of the He-shell so 
that the heating induced by the energy delivered by this chain does not affects the 
point where the He-flash is ignited.

To investigate more in detail this point we have computed an additional set of models 
in which He-rich matter (Y=0.98 and Z=0.02) is accreted onto a cooled down CO WD at 
rates suitable for an He detonation ($\dot{M}=1\div 4\times 10^{-8}$\msun ).
The initial model for the CO WD has been obtained by evolving a 0.6 \msun pure He star 
(Y=0.98, Z=0.02) from the He Main Sequence down to the cooling sequence. The luminosity, 
the effective temperature, the central temperature, and density of this structure are 
the following: $\log(L/L_{\odot})=-1.905,\ \log(T_e)=4.236,\ \log(T_c)=7.310\ 
{\rm and}\ \log(\rho_c)=6.585$. 
A glance at the data plotted in Fig. 6 shows that at the physical base of the 
He-layers the \n14 abundance is zero. The main physical parameters at the onset of 
the He-flash are listed for each model in Table 1: the first column gives the 
accretion rate (\msun\yrm1) while the others give the thickness in mass of the 
He-layers (solar units), the temperature (K), the density ($g/cm^3$), and the 
degeneration parameter at the point where the energy production via He-burning attains 
the maximum value. The quantities referred to the case which neglects the NCO energy 
contribution are listed in columns 2 to 5, while those referred to the case which 
includes the NCO reactions are listed in columns 6 to 10. 

\placefigure{fig6}

\placefigure{fig7}

\placetable{table1}

Fig. 7 shows the evolution in the $\rho - T$ plane of the bottom of the He layer for 
models computed at different $\dot{M}$ and neglecting the NCO energy contribution. 
Note that only for accretion rates smaller than $3\times 10^{-8}$ \msun\yrm1 the NCO 
chain could become active. In fact, only for these values of $\dot{M}$ the density at 
the physical base of the He-shell exceeds $\rho_{th}$ (dashed line). 

Fig. 8 depicts the evolution in the $\rho - T$ plane of the base of the He-shell with 
(dashed line) and without (solid line) the inclusion of NCO reaction in the nuclear 
network. The evolution of the mass point where the energy production is at maximum is 
plotted in Fig. 9. As soon as the density becomes greater than $\rho_{th}$, the NCO 
reaction is ignited and releases a small amount of energy. The continuous accretion 
process causes an increase in the shell density; therefore during the subsequent 
evolutions the physical conditions suitable for the NCO reaction are reached in more 
and more external layers. As a matter of fact, a \c14 burning shell forms and moves 
outward in mass all over the accretion process up to the onset of the He-flash. 

\placefigure{fig8}

Interestingly enough, an increase in the accretion rate causes a rapid outward shift 
of the \c14 shell. This occurrence is simply explained by noting that the He-shell is 
globally less degenerate due to the increase in the accretion rate. Data plotted in 
Fig. 9, and the physical parameters listed in Table 1, clearly support the evidence 
that the effects of NCO chain are more relevant at intermediate accretion rates 
($\dot{M}=1.5\div 2 \times 10^{-8}$ \msun\yrm1). In fact, at lower accretion rates the 
He-shell is more degenerate and the \c14 shell detaches more and more from the bottom 
of the He-shell. On the contrary, for higher accretion rates the physical conditions 
suitable for the electron capture are attained when \3a reaction already produces a 
huge amount of energy. Even at higher accretion rates (higher than $3\times10^{-8}$ 
\msun\yrm1) the density of the He-shell does never exceed the critical density 
$\rho_{th}$, and therefore the NCO reaction does not supply any energy contribution.

This notwithstanding the inclusion of NCO energy contribution does not change the 
final fate of these structures, all of them should undergo an explosive event.

\placefigure{fig9}

\placetable{table2}

\section{Final remarks}

We have shown that the effect of NCO reaction on the onset of He-burning in degenerate 
physical conditions is negligible. In fact, for low-mass structures climbing the RGB, 
the models with high metallicity (Z=0.001 and Z=0.02) do not attain densities so high 
to ignite the NCO reaction,while more metal-poor models present such a low abundance 
of \n14 that the energy released by $\alpha$-captures on \c14 does not affect the 
evolution. Of course if the $\rho_{th}$ value was smaller than we assumed then the 
energy contribution of the NCO chain should be reevaluated. 
Accounting for this evidence, we have decided to perform several numerical experiments 
for a $0.8M_\odot$ solar metallicity model, arbitrarily decreasing the value of 
$\rho_{th}$ in order to test to what extend present results are affected by the 
uncertainty on this parameter, as well as to verify if stellar observables allow us 
to put a firm constraint on this quantity. From the data listed in Table 2, 
it is possible to notice that a decrease of the order of 20\% of 
$\rho_{th}$ has no effect on the evolutionary properties of the model. However, when 
the value of  $\rho_{th}$ is decreased by a larger factor, the numerical experiments 
clearly reveal that the size of the He core at the He flash onset and, in turn, the 
luminosity of the RGB tip and of the ZAHB are significantly affected by the NCO energy 
release. In particular, for a decrease of 30\% of $\rho_{th}$, we obtain an RGB tip 
luminosity of $\log(L_{tip}/L_{\odot})=3.362$ and an He core mass of $M_{He}=0.4612$ 
\msun to be compared with the values of the standard case, namely 
$\log(L_{tip}/L_{\odot})=3.430$ and $M_{He}=0.4753$ \msunend. From an observational
point of view, we can estimate that this occurrence produces an increase in the
visual magnitude of the RGB Tip and of the ZAHB equal to $\approx 0.2$ mag and 
$\approx 0.1$ mag, respectively.
These values are well within the current observational and theoretical uncertainties 
which still affect these crucial observables (Cassisi et al. 1998, Salaris \& 
Cassisi 1998). However, a further 
decrease ($\approx40$\%) in the value of $\rho_{th}$ (see data in Table 2) produces 
quite larger effects. These changes could be easily tested, allowing to put a firm 
constraint on $\rho_{th}$ value, once the sample of high-metallicity clusters with 
reliable RGB Tip luminosity estimates will be increased.

Following the suggestion of an anonymous referee we also explored the dependence of 
our results on the value of the ${^{14}C}(\alpha,\gamma){^{18}O}$ cross section.
The current scenario (Funck \& Langanke 1989; Goerres et al. 1992) states that at low 
temperatures ($T < 10^8$ K) the cross section for this reaction rate is dominated by a 
direct capture contribution which is affected by an uncertainty of roughly two orders 
of magnitude. Owing to this indisputable fact, we performed several numerical 
experiments by arbitrarily changing the value of the cross section for the $\alpha$ 
capture on the \c14. In particular, the value of $N_A <\sigma v>_{^{14}C,\alpha}$ 
by Caughlan 
\& Fowler (1988) has been multiplied by factors equal to $10^{-2}$, $10^{-1}$, 
$10^{1}$ and $10^{2}$. 

\placefigure{fig10}

\placetable{table3}

As far as the $M=0.6 M_{\odot}$ model at $Z=10^{-4}$ is 
concerned (see \S 3), Fig. 10 shows the evolutions of the center in the $\rho - T$ 
plane. Data plotted in this figure suggest that an increase in the reaction rate of 
${^{14}C}(\alpha,\gamma){^{18}O}$ does not cause a significant change in the canonical 
scenario, except for a slightly faster release of the energy. 
In particular, it is worth noticing that the profiles associated with an increase of 
one and two order of magnitude overlap, respectively. This occurrence is a direct 
consequence of the fact that the efficiency of the NCO cycle is only triggered by the 
e-capture on \n14. On the other hand, a decrease in the reaction rate of the $\alpha$ 
capture on the \c14, causes a decrease in the energy released, and in turn an increase 
in its duration up to the onset of the He-flash. In any case the He-core mass and the 
luminosity level of the Tip remain unchanged when compared with the standard case. 

The same experiments were also performed for the model at $M=0.8 M_{\odot}$ and 
$Z=0.02$ by assuming a decrease of the order of 30\% in the transition density 
for the e-capture. Tab. 3 
lists the values of the He core mass, central temperature, and density at the onset 
of the He-flash. Once again the increase in the reaction rate of the $\alpha$ capture 
on the \c14 does not cause a significant change of the physical conditions at the 
onset of the He-flash. On the contrary, a decrease in the value of 
$N_A <\sigma v>_{^{14}C,\alpha}$ causes a decrease in the degeneration of the He core, 
and therefore a mild decrease in the He core mass at the onset of the He flash. 
Unfortunately, the variations in the observables (luminosity level of tip and 
Horizontal Branch) are too small to constrain the efficiency of this reaction.

In the case of typical mass CO WDs accreting He-rich matter we showed that the final 
outcome depends only on the accretion rate, while the NCO reaction only causes a 
slight decrease in the He layers mass at the He-flash. This occurrence implies that 
the inclusion of the NCO reaction cannot switch an explosive He-burning into a 
quiescent one in such a way that the CO core can grow in mass up to the Chandrasekhar 
mass limit. In fact, for low $\dot{M}$ values the energy contribution given by the NCO 
reaction is not able to heat up the bottom of the He-shell, while for higher accretion 
rates the physical conditions for the e-capture are never reached.

These considerations are still valid when He-rich matter is accreted as a by-product 
of a H-burning shell. In fact, for higher accretion rates (higher than $4\times 
10^{-8}$ \msun\yrm1) the presence of the H-burning shell increases the temperature of 
underlying He-shell, thus preventing it to become degenerate. Therefore the physical 
conditions for the ignition of NCO chain are never attained (Cassisi, Iben \& 
Tornamb\'e 1998). On the contrary, for lower accretion rates He and H shells decouple 
and the behavior of the models accreting H-rich matter are identical to the models 
accreting He directly (Piersanti et al.2000).

\acknowledgments{We wish to warmly thank G. Bono for several suggestions and useful
discussion on this topic, as well as for the generous and painstaking effort provided 
for improving the readability of this paper. Useful comments and suggestions from an 
anonymous referee are also acknowledged. This work was supported by MURST (Cofin2000) 
under the scientific project "Stellar observables of cosmological relevance".
}


\newpage
\figcaption[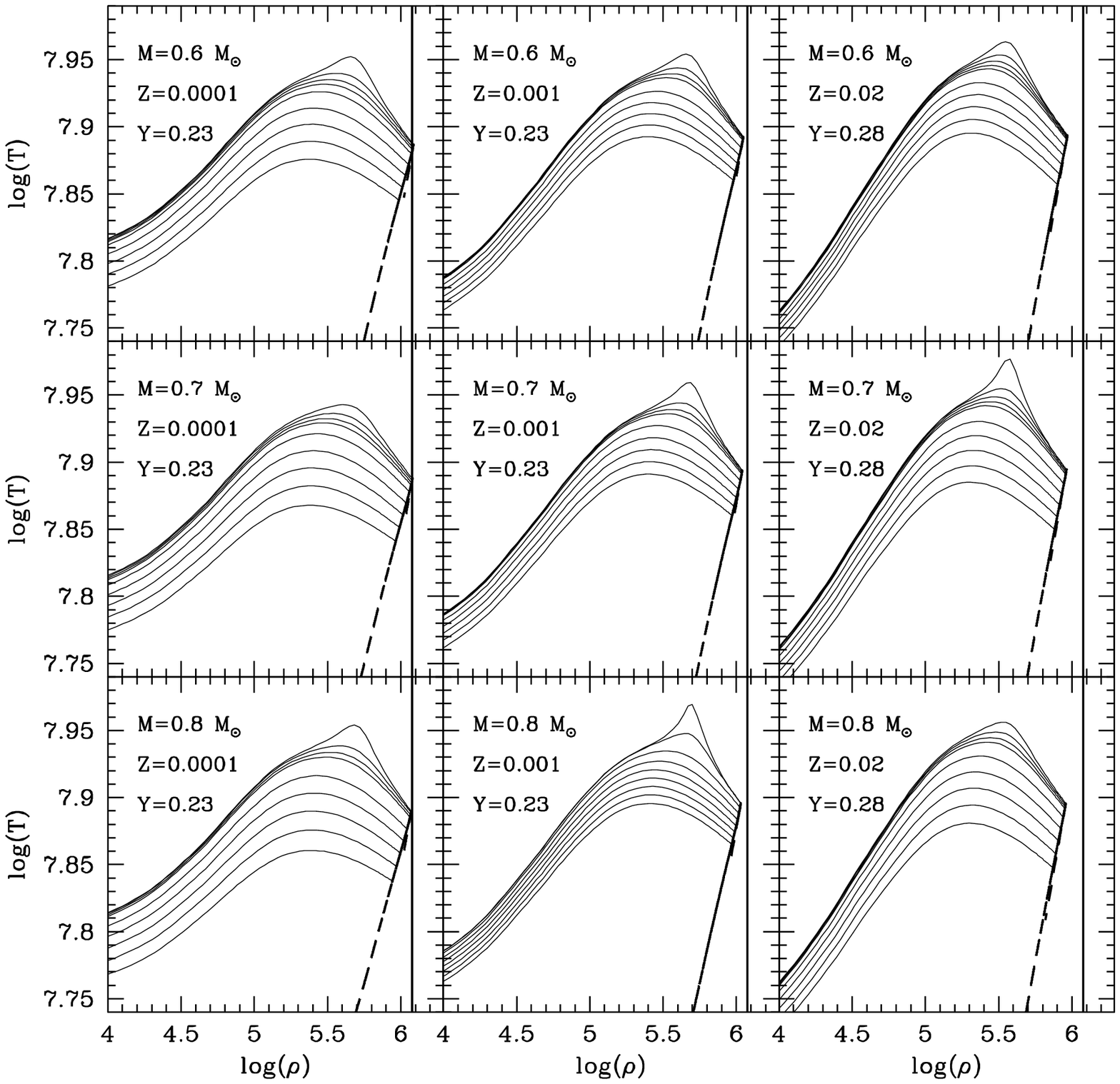] {Profiles in the $\rho -T$ plane of stars climbing 
the RGB up to the onset of the central He-flash. For each evolving mass 
of fixed metallicity (see labeled values) physical properties at 
different luminosity levels are plotted. The model at the He ignition 
is the top line. The dashed line shows the evolution as a function of 
time of the physical conditions at the center. The
heavy solid line on the right of each panel marks the region suitable for
electron capture to occur: for structures located at the right of this 
line e-captures occurs. 
\label{fig1}}

\figcaption[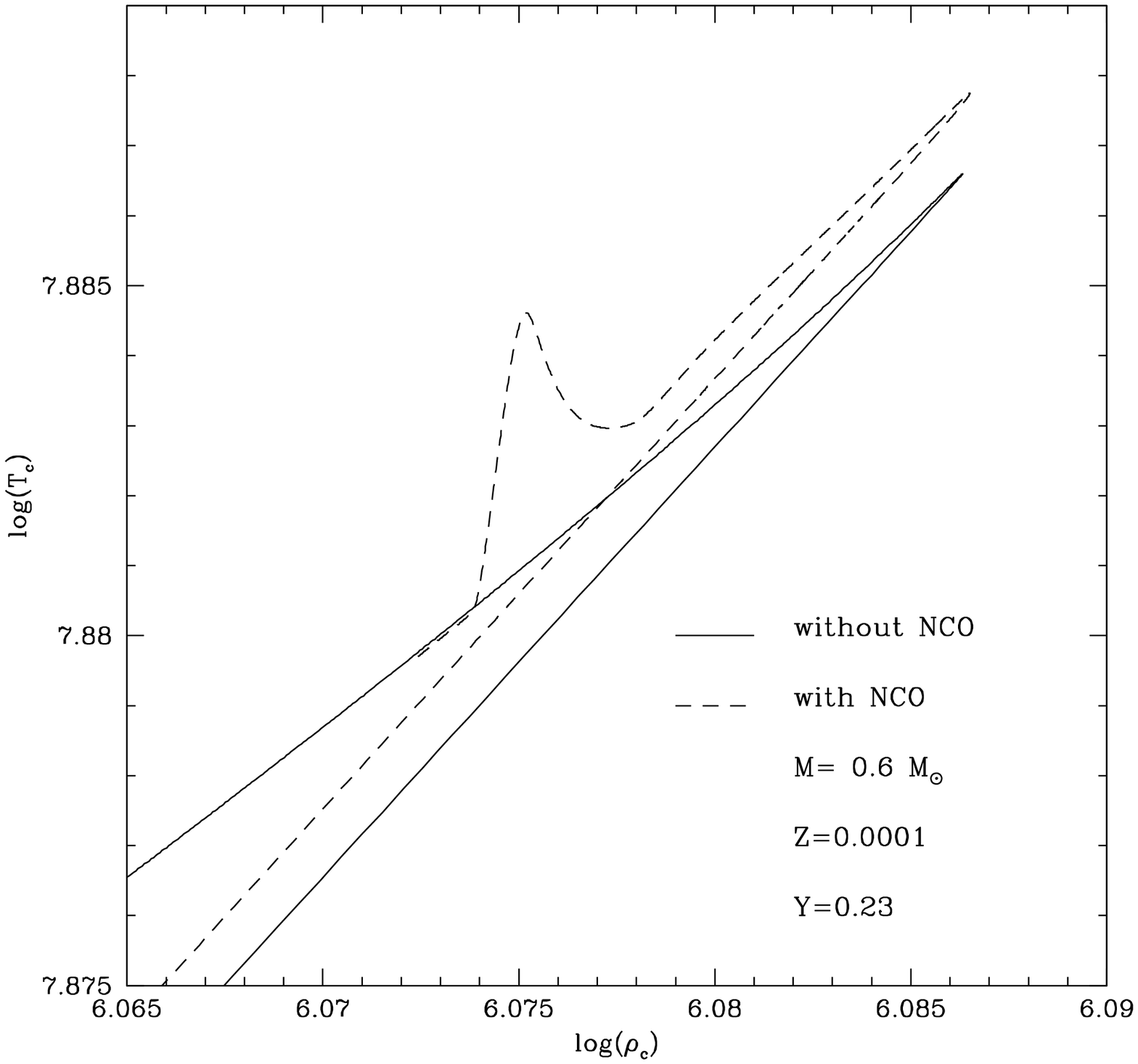] {The evolution in the $\rho - T$ plane of the 
center of the model with M=0.6 \msun and Z=0.0001. The solid lines 
refer to the model in which the NCO energy contribution is not taken 
into account, while the dashed one refers to the model which includes
of the energy contribution given by NCO reaction.
\label{fig2}}

\figcaption[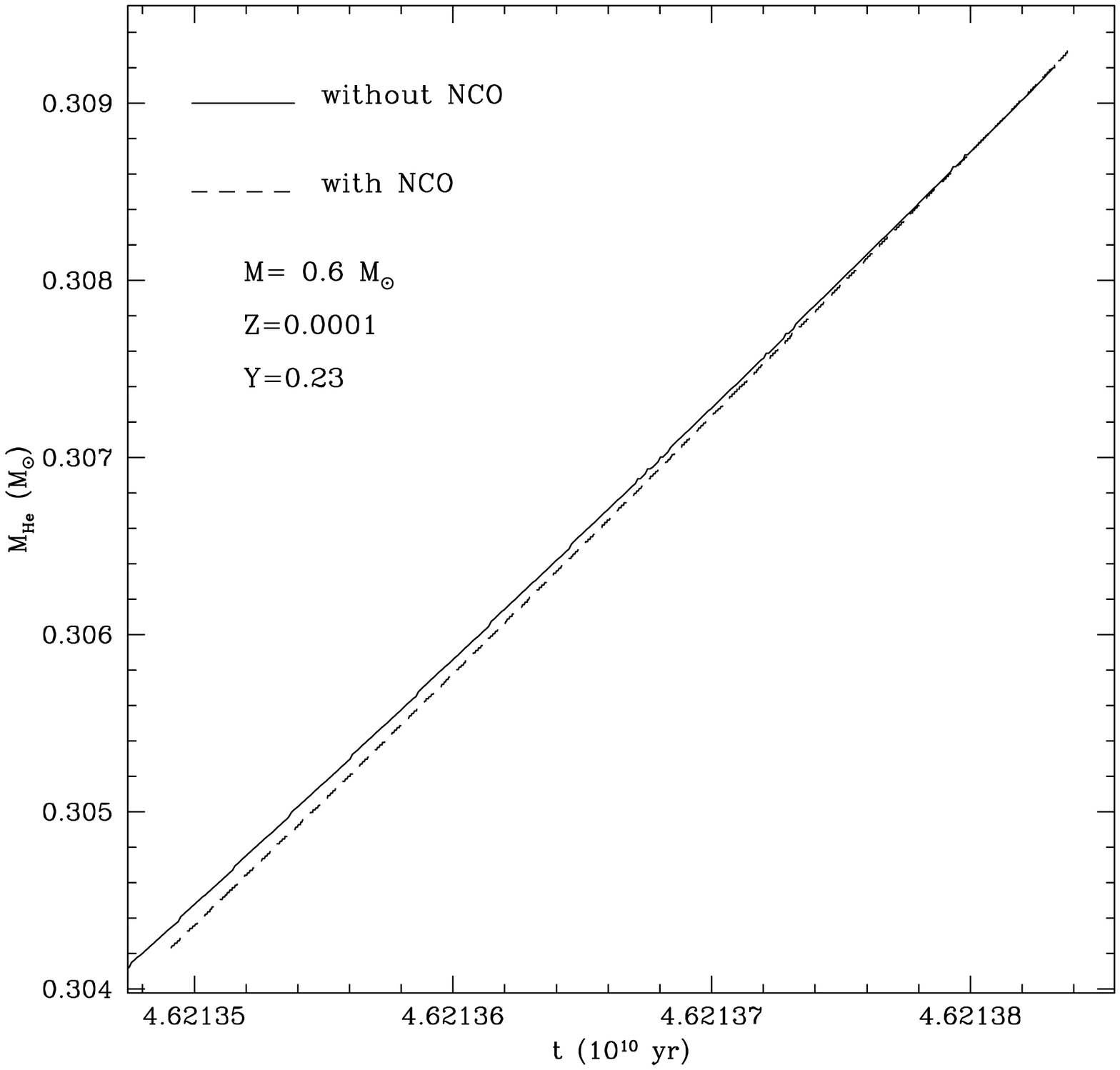] {Time evolution of the He-core mass for the model 
at M=0.6 \msun and Z=0.0001. Dashed and solid lines refer to models 
in which the NCO energy contribution has been included and neglected 
respectively. 
\label{fig3}}

\figcaption[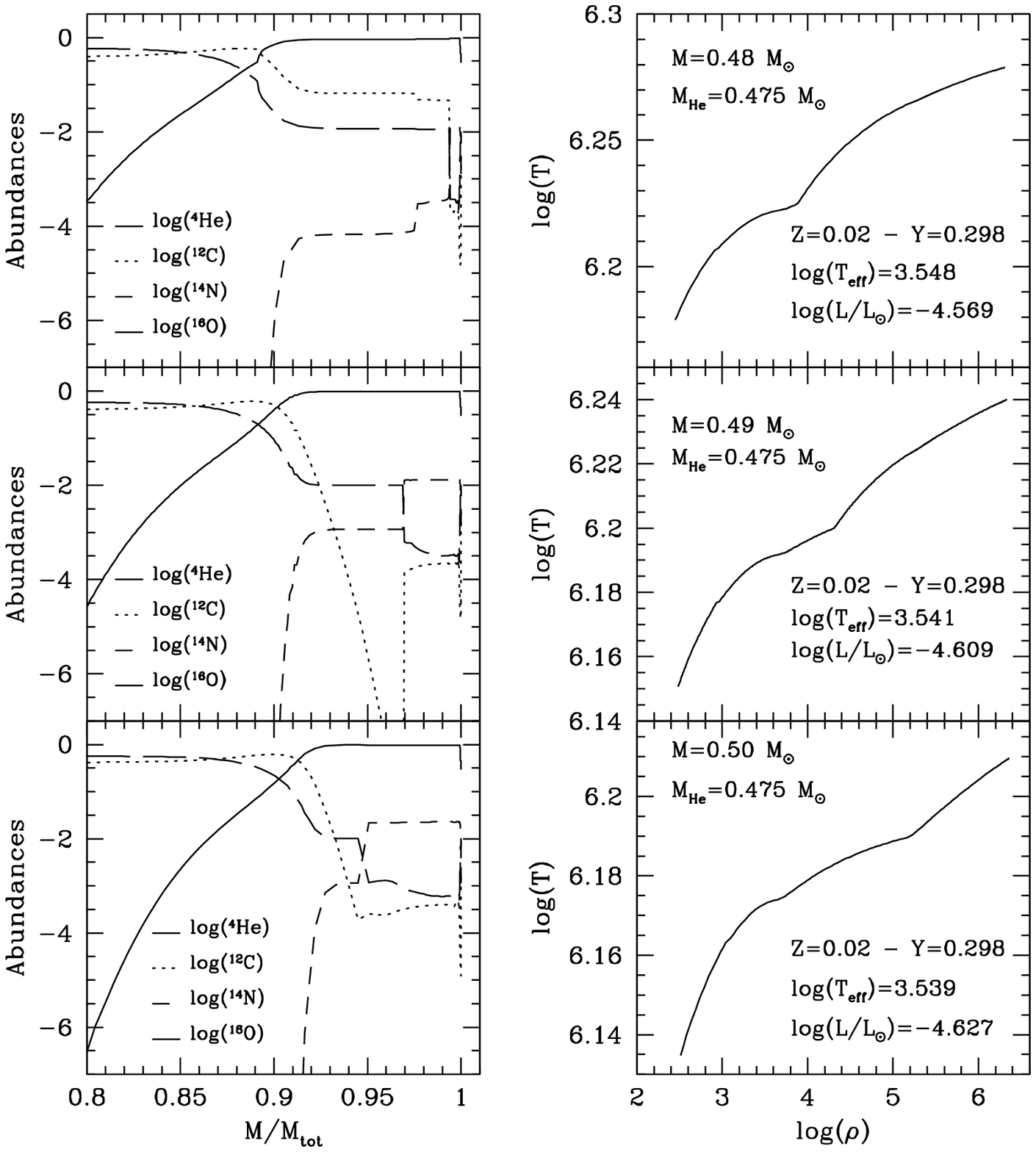] {Chemical abundances in the most external layers 
of the He-shell (left panels) and profiles in the $\rho - T$ plane 
(right panels) for models of CO WDs obtained evolving structures 
with Z=0.02, Y=0.298, mass of the He-core of 0.475 \msun, and different 
total masses (see labeled values) from the Zero Age Horizontal Branch 
down to the cooling sequence. The values of luminosity, effective 
temperature, central temperature, and density for each model are 
plotted in the right panels. In the left panels the solid line refers 
to $\log({^{4}He})$, the dotted one to $\log({^{12}C})$, 
the short-dashed one to $\log({^{14}N})$, and the long-dashed one to 
$\log({^{16}O})$.
\label{fig4}}

\figcaption[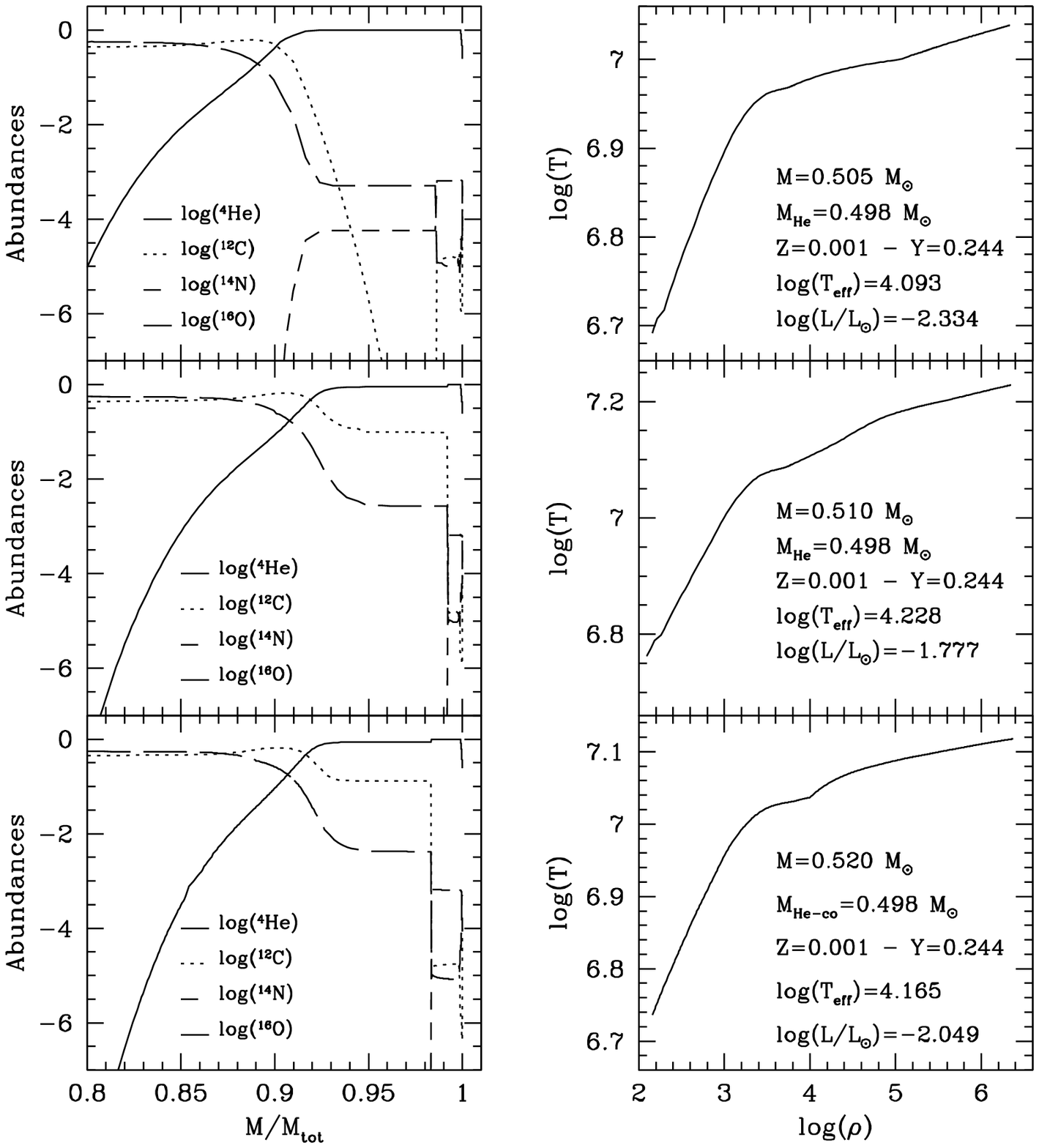] {The same as in Fig. 4, but for the case Z=0.001 
and Y=0.244.
\label{fig5}}

\figcaption[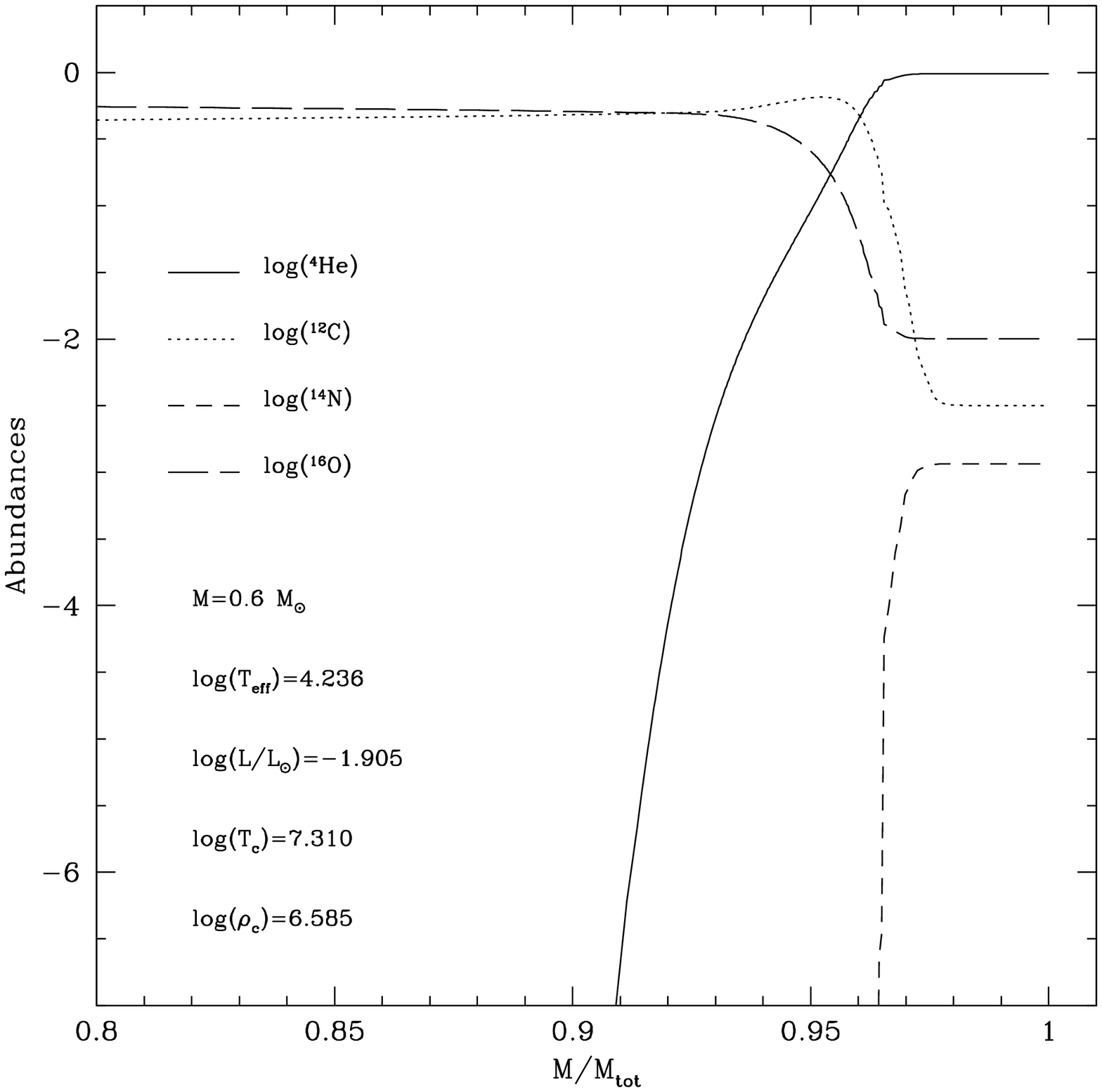] {Chemical abundances in the most external layers 
of a 0.6 \msun CO WD (see text for more details). 
\label{fig6}}

\figcaption[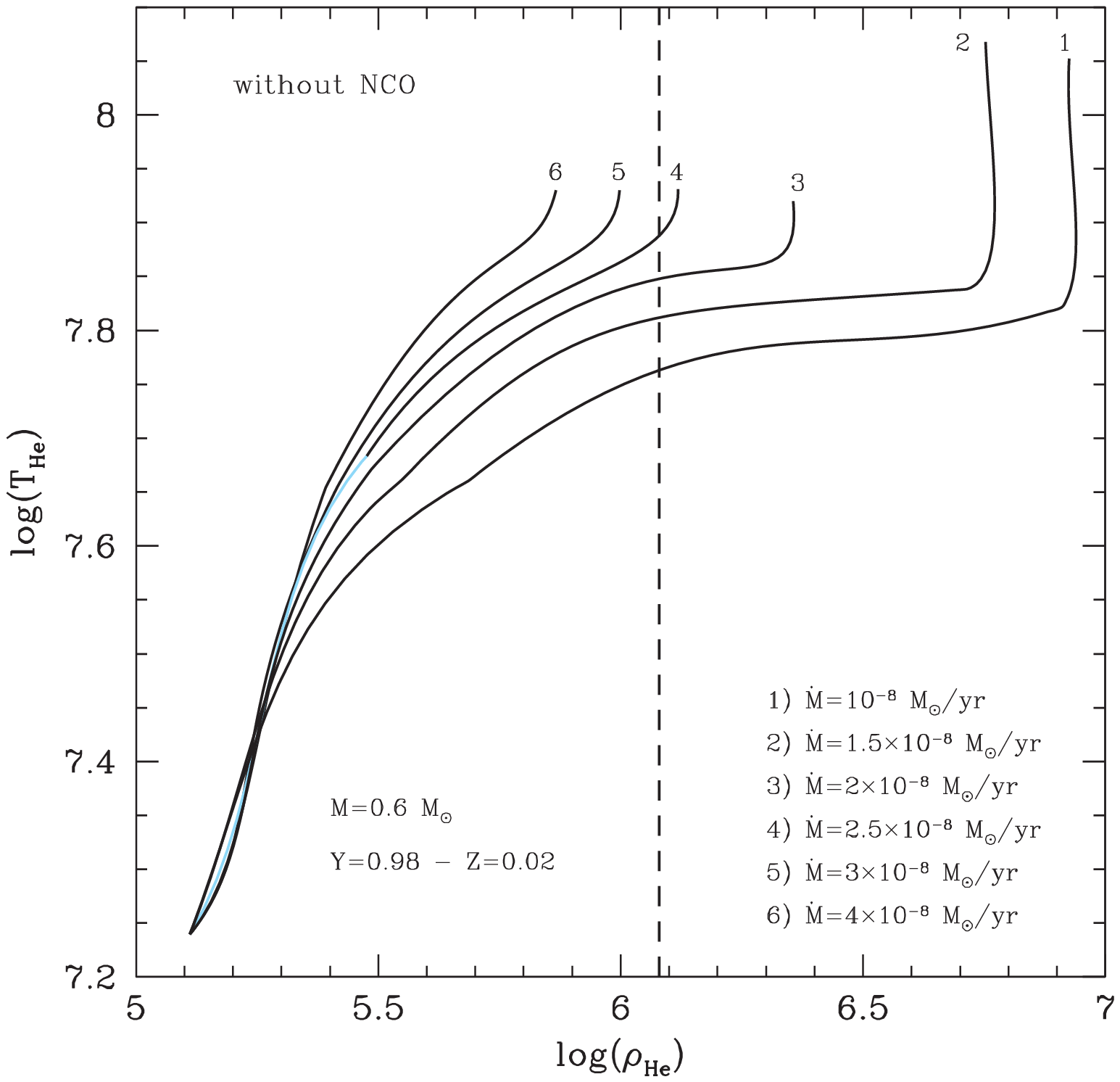] {The evolution in the $\rho - T$ plane of the 
physical base of the
He-shell for models accreting He-rich matter with solar metallicity at 
different accretion rates (as labelled) onto a cooled down CO WD of 
0.6 \msunend .
\label{fig7}}

\figcaption[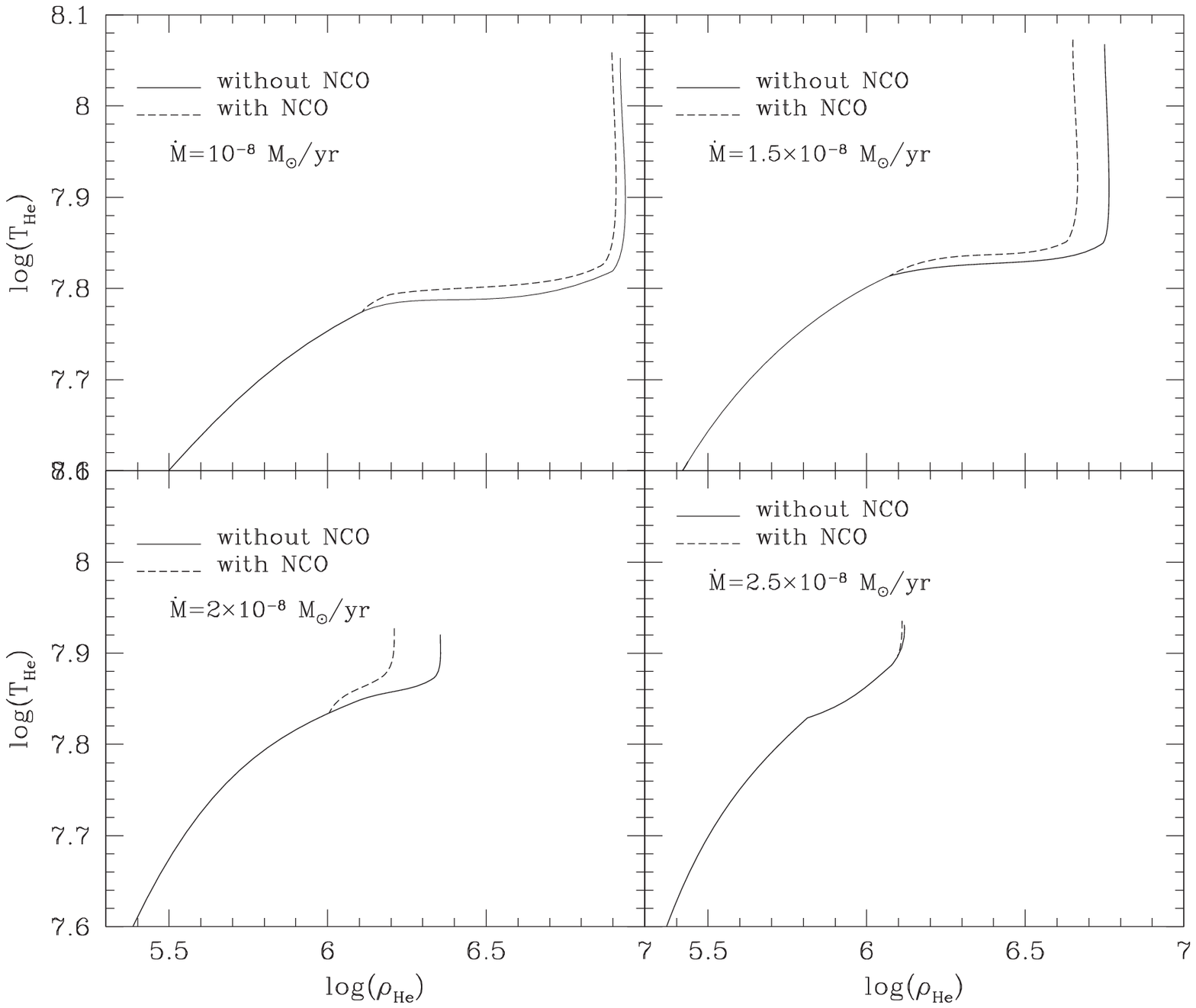] {The evolution in the $\rho - T$ plane of the 
physical base of the
He-shell for models accreting He-rich matter at different accretion rates 
(as labelled). The solid lines refer to the case in which the NCO
energy contribution is not take into account, while the dashed ones 
to the case in which NCO reaction has been included in the nuclear 
network.
\label{fig8}}

\figcaption[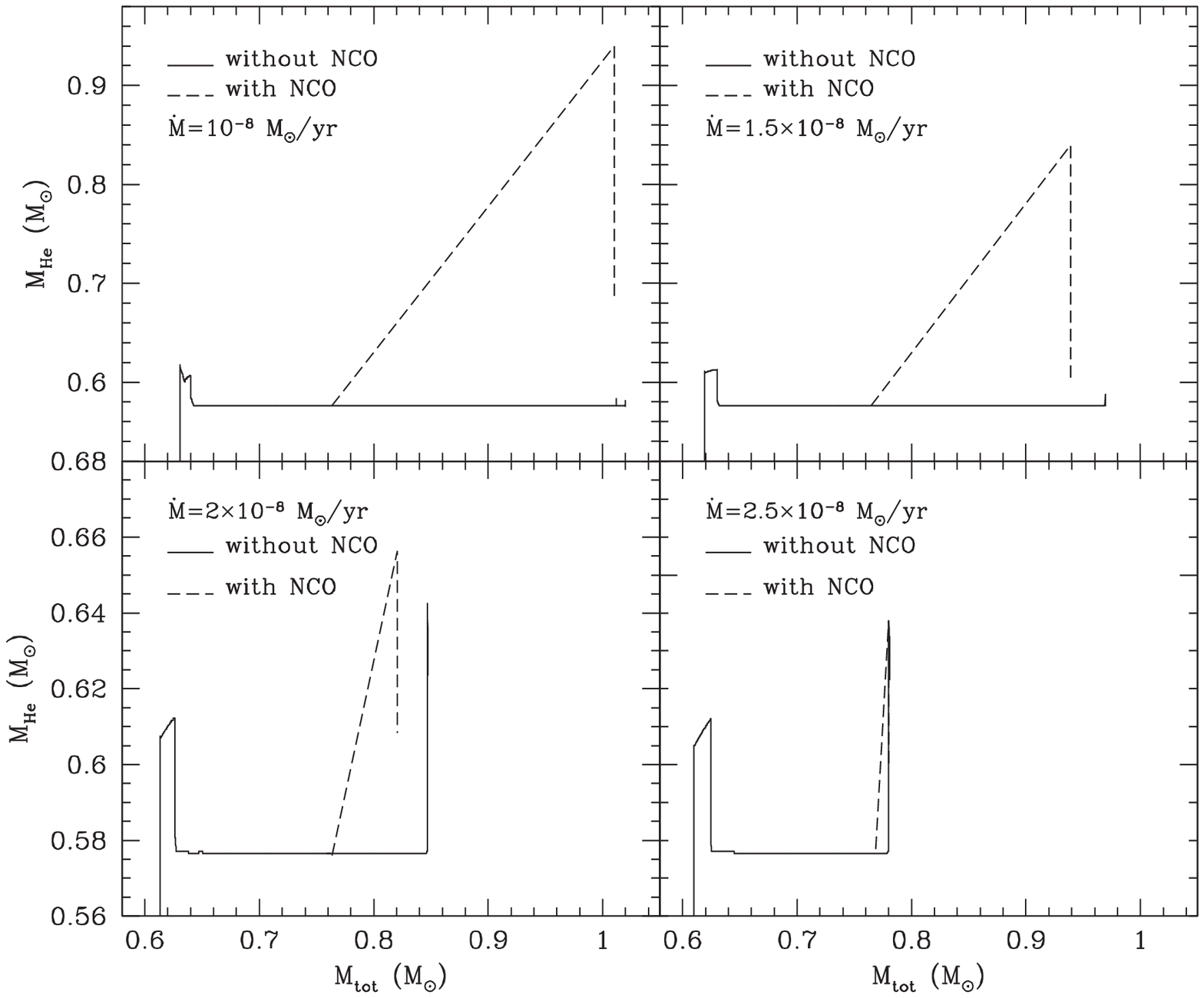] {The evolution of the mass point where the energy production is
maximum as function of time. The dashed and solid lines show  
models in which NCO energy contribution is included and neglected 
respectively. 
\label{fig9}}

\figcaption[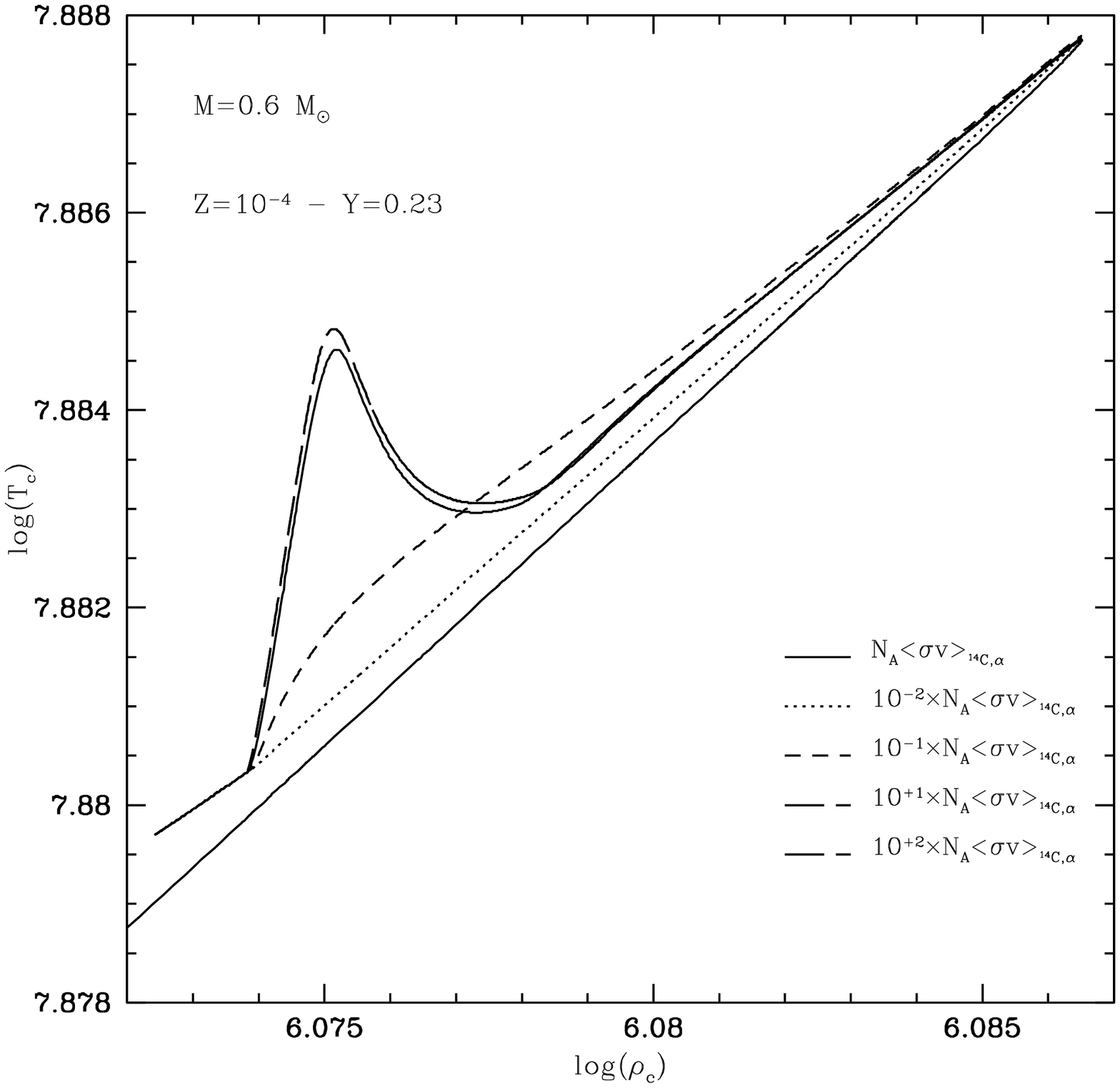] {Evolution in the $\rho - T$ plane of the 
center of the model with M=0.6 \msun and Z=0.0001 in which the energy 
contribution given by NCO reaction has been taken into account. 
Individual lines refer to different values adopted for the 
${^{14}C}(\alpha,\gamma){^{18}O}$ cross section as labelled inside 
the figure (see text).
\label{fig10}}


\newpage


\begin{deluxetable}{rcccccccc}
\tablecaption{Accretion of He-rich matter onto a CO WD of M=0.6 \msun .
\label{table1}
}
\tablehead{
\colhead{$\dot{M}$}                  &
\colhead{$\Delta M_{He}$}            &
\colhead{$\log(T_{He})$}             &
\colhead{$\log(\rho_{He})$}          &
\colhead{$\psi$}                     &
\colhead{$\Delta M_{He}$}            &
\colhead{$\log(T_{He})$}             &
\colhead{$\log(\rho_{He})$}          &
\colhead{$\psi$}                     
}
\startdata
 & \multicolumn{4}{c}{without NCO} & \multicolumn{4}{c}{with NCO} \nl
\hrulefill & \multicolumn{4}{c}\hrulefill & \multicolumn{4}{c}\hrulefill \nl
$10^{-8}$           & 0.443 & 7.877 & 6.932 & 75 & 0.436 & 7.883 & 6.905 & 71 \nl
$1.5\times 10^{-8}$ & 0.393 & 7.895 & 6.742 & 57 & 0.363 & 7.903 & 6.648 & 50 \nl
$2\times 10^{-8}$   & 0.271 & 7.933 & 6.230 & 27 & 0.244 & 7.932 & 6.163 & 25 \nl
$2.5\times 10^{-8}$ & 0.204 & 7.950 & 5.965 & 18 & 0.203 & 7.953 & 6.008 & 19 \nl
$3\times 10^{-8}$   & 0.174 & 7.966 & 5.805 & 14 & -    &   -   &   -    &  - \nl
$4\times 10^{-8}$   & 0.142 & 7.994 & 5.623 & 10 & -    &   -   &   -    &  - \nl
\enddata
\nl
\end{deluxetable}


\begin{deluxetable}{ccc}
\tablecaption{The effect of different reduction factor of $\rho_{th}$ on the evolution 
of a 0.8 \msun Z=0.02 model.
\label{table2}
}
\tablehead{
\colhead{$\Delta\rho_{th}\over\rho_{th}$}       &
\colhead{$M_{He}$ (\msun)}          &
\colhead{$\log(L/L_\odot)^{tip}$}
}
\startdata
0.00 & 0.4753 & 3.430 \nl
0.20 & 0.4753 & 3.430 \nl
0.25 & 0.4730 & 3.424 \nl
0.30 & 0.4612 & 3.362 \nl
0.35 & 0.4472 & 3.304 \nl
0.40 & 0.4375 & 3.255 \nl
0.45 & 0.4308 & 3.219 \nl
0.50 & 0.4260 & 3.192 \nl
\enddata
\nl
\end{deluxetable}


\begin{deluxetable}{lccc}
\tablecaption{Effect of different value for the ${^{14}C}(\alpha,\gamma){^{18}O}$ reaction rate
on the evolution of a 0.8 \msun Z=0.02 model.
\label{table3}
}
\tablehead{
\colhead{$N_A <\sigma v>_{^{14}C,\alpha}$} &
\colhead{$M_{He}$ (\msun)}                 &
\colhead{$\log(T_{c})$}                    &
\colhead{$\log(\rho_{c})$}
}
\startdata
CF88 $\times 10^{-2}$ & 0.4604 & 8.0598 & 5.8931 \nl
CF88 $\times 10^{-1}$ & 0.4612 & 8.0599 & 5.8953 \nl
CF88                  & 0.4612 & 8.0577 & 5.8964 \nl
CF88 $\times 10$      & 0.4612 & 8.0577 & 5.8965 \nl
CF88 $\times 10^{2}$  & 0.4613 & 8.0584 & 5.8962 \nl
\enddata
\nl
\end{deluxetable}

\end{document}